\documentclass[journal]{IEEEtran}

\usepackage{ucs}
\usepackage[utf8x]{inputenc}
\usepackage[cmex10]{amsmath}
\usepackage{cite, amsfonts, amssymb, amsthm, bm, bbm, graphicx, relsize, multirow, booktabs, tikz,subfigure,soul}
\usepackage[american]{babel}
\usepackage[T1]{fontenc}
\usepackage{algorithmic, algorithm}
\usepackage[multiple]{footmisc}
\usepackage{microtype}
\theoremstyle{definition}

\theoremstyle{remark}

\begin{document}

\title{On Clustered Statistical MIMO Millimeter Wave Channel Simulation}
\author{Stefano Buzzi, {\em Senior Member}, {\em IEEE}, and Carmen D'Andrea
\thanks{The authors are with the Department of Electrical and Information Engineering, University of Cassino and Lazio Meridionale, I-03043 Cassino, Italy (buzzi@unicas.it, carmen.dandrea@unicas.it).}
}
\maketitle

\begin{abstract}
The use of mmWave frequencies  is one of the key strategies to achieve the fascinating 1000x increase in the capacity of future 5G wireless systems. 
While for traditional sub-6 GHz cellular frequencies several well-developed statistical channel models are available for system simulation, similar tools are not available for mmWave frequencies, thus preventing a fair comparison of independently developed transmission and reception schemes. 
In this paper we provide a simple albeit accurate statistical procedure for the generation of a clustered MIMO channel model operating at mmWaves, for both the cases of slowly and rapidly time-varying channels. Matlab scripts for channel generation are also provided, along with an example of their use.

\end{abstract}

\section{Introduction}
The use of millimeter-wave (mmWave) frequencies  for cellular communications 
is one of the key technologies for greatly improving the capacity of future wireless networks \cite{whatwillbe}. 
While being unsuited for macro-cellular communications, recent measurements have shown that mmWave
frequencies work fine for cellular communications over short distances (up to 100-200 meters at most) \cite{ghoshmmwave,itwillwork,baldemair2015ultra}. 

\subsection{Related work}  
Based on the evidence that propagation mechanisms at mmWave frequencies are totally different from those at sub-6 GHz bands,  cellular channel modeling and validation for mmWave frequencies has been an intense research track in the last few years. The paper \cite{PrecodingHeath} is one of the first to propose a clustered channel model for cellular systems operating at mmWave frequencies; no multipath and, mostly important, no path-loss model are considered, however, and the results are presented in terms of received SNR, with no insight on what performance to expect for given values of the transmit power and of the link length. 
The paper \cite{FrequencySelectiveHeath} introduces a clustered channel model including the path loss and multipath;  the multipath delay spread here arises from the chosen length of the cyclic prefix (whereas in practical situations exactly the opposite happens), and  all the paths are assumed to have the same average power.
In \cite{rappaport2014},  a narrowband clustered time-varying channel model is proposed, with the number of clusters modeled as a Poisson-distributed random variate and with a fixed number of sub-paths for each cluster.
The paper \cite{rappaport2014Wideband} provides parameters for the wideband clustered millimeter wave channel, such as the number of clusters, the number of cluster sub-paths, the delays of each cluster and the direction of arrival and departures. Nevertheless, it is not explained how these values can be assembled together to obtain a matrix-valued discrete-time channel impulse response.
Three path loss models and several parameters  (the fixed number of clusters and the fixed number of sub-paths for each cluster, the exponential distribution for the delays, the Gaussian distribution for azimuth arrival and departures angles and the Laplacian distribution for elevation arrival and departures angles) for the mmWave channel are detailed in \cite{rappaport2016channel}, but no procedure for using these parameters in order to have a matrix-valued channel impulse response is given, neither the effect of transmit and receive pulse shapes is considered.
Accurate clustered channel models are proposed also in references \cite{METIS2015, MiWEBA2014,COSTIC1004}, which are deliverables of EU-funded research projects, but the channel simulators, although accurately described, seemingly are not publicly available. A geometry-based stochastic channel model is then proposed in the framework of the COST 2100 Action \cite{COST2100} for sub-6 GHz frequencies.  

More recently,
the papers \cite{rappaport2015channel, rappaport2016MIMOchannel} have provided an elegant  continuous-time representation of the channel impulse response at mmWave  using time cluster-spatial lobes parameters; the papers, however,  do not take into account the time-variance of the channel and do not provide a procedure for obtaining a matrix-valued sequence describing the MIMO channel impulse response, also as a function of the transmit and receive pulse shapes. 
\subsection{Paper contribution}
This letter builds upon previously considered models and provides a clustered statistical MIMO channel model with \textit{all} the features that have been only partially considered in previous references. In particular, the proposed model 
accounts for  path loss, angular diversity, random number of clusters, possible line-of-sight (LOS) link, frequency-selectivity, contribution from the transmit and receive shaping filters, and, finally, time-variance of the channel. Matlab scripts for channel simulation are also given \cite{github_dandrea}, to enable reproducible research and fast verification of transceiver processing algorithms. 

The remainder of this paper is organized as follows. Next Section  provides the channel model for the case of static channel, while the generation of a time-varying channel is addressed in Section III. Section IV shows a simple application of the proposed channel model, while, finally, concluding remarks are given in Section V.

\textit{Notation:} Uppercase boldface letters (i.e., $\mathbf{A}$) denote a matrix; lowercase boldface letters (i.e., $\mathbf{a}$) denote a vector; $(\cdot)^H$ denotes conjugate transpose. $ \left\|{\cdot}\right\|_{\rm F}$ denotes the Frobenius norm; statistical expectation is denoted by $E[\cdot]$, while $\mathcal{CN}(\mu,\sigma^2)$ is a complex Gaussian random variable with mean $\mu$ and variance $\sigma^2$; $\mathcal{U}(a ,b)$ is a  random variable uniformly distributed in $[a,b]$.

\section{Time-invariant channel model}
We focus on a system with $N_T$ transmit antennas and $N_R$ receive antennas, and start by considering the case of slow fading, i.e. the channel coherence time exceeds the observation time. 
According to the clustered model, we assume that the propagation environment is made of $N_{\rm cl}$ scattering clusters, each of which contributes with $N_{{\rm ray}, i}$ propagation paths  $i=1, \ldots, N_{cl}$, plus a 
possibly present LOS component. 
We denote by  $\phi_{i,l}^r$ and $\phi_{i,l}^t$ the azimuth angles of arrival and departure of the $l^{th}$ ray in the $i^{th}$ scattering cluster, respectively; similarly, $\theta_{i,l}^r$ and $\theta_{i,l}^t$ are the elevation angles  of arrival and departure of the $l^{th}$ ray in the $i^{th}$ scattering cluster, respectively. We also include in the model the heights at which the transmit and receive antennas are located, so as to take into account ground surface reflection. 
\begin{figure}[!h]
\centering
\includegraphics[scale=0.3]{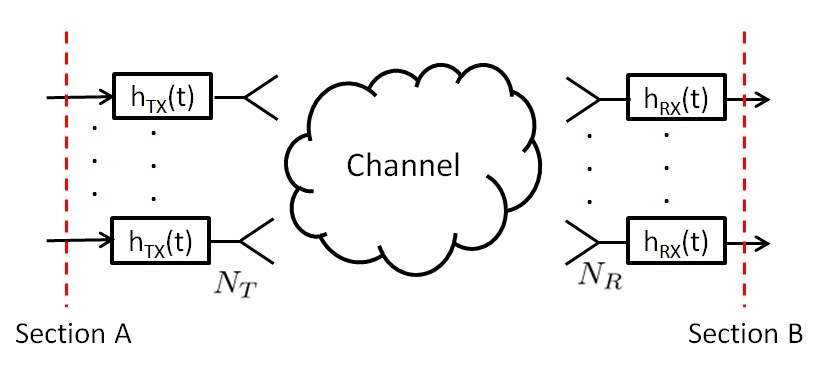}
\caption{The input and output sections of the modeled system.}
\label{Fig:Channel}
\end{figure}
Denoting by $h_{\rm TX}(t)$ the baseband equivalent of the transmitted waveform, by 
$h_{\rm RX}(t)$ the baseband equivalent of the impulse response of the receive filter, and by $h(t)=h_{\rm TX}(t) \ast h_{\rm RX}(t)$ their convolution, 
the impulse-response of the linear time-invariant system between section A and section B in Fig. \ref{Fig:Channel}  is a matrix-valued (of dimension $N_R \times N_T$) time-continuous function that  can be written as follows:
\begin{multline}
\mathbf{H}(\tau)=\gamma\sum_{i=1}^{N_{\rm cl}}\sum_{l=1}^{N_{{\rm ray},i}}\alpha_{i,l}
\sqrt{L(r_{i,l})} \mathbf{a}_r(\phi_{i,l}^r,\theta_{i,l}^r)\times \\ \mathbf{a}_t^H(\phi_{i,l}^t,\theta_{i,l}^t)  h(\tau-\tau_{i,l}) + \mathbf{H}_{\rm LOS}(\tau)\; .
\label{eq:channel1}
\end{multline}
In the above equation, neglecting for the moment $\mathbf{H}_{\rm LOS}(\cdot)$, to be specified later, $\alpha_{i,l}$ and $L(r_{i,l})$ are the complex path gain and the attenuation associated  to the $(i,l)$-th propagation path (whose length is denoted by $r_{i,l}$),  respectively; $\tau_{i,l}=r_{i,l}/c$, with $c$ the speed of light,  is the propagation delay associated with the $(i,l)$-th path. 
The complex gain  $\alpha_{i,l}\thicksim \mathcal{CN}(0, \sigma_{\alpha,i}^2)$, with  $\sigma_{\alpha,i}^2=1$\cite{PrecodingHeath}. The factors $\mathbf{a}_r(\phi_{i,l}^r,\theta_{i,l}^r)$ and $\mathbf{a}_t(\phi_{i,l}^t,\theta_{i,l}^t)$ represent the normalized receive and transmit array response vectors evaluated at the corresponding angles of arrival and departure; finally, $\gamma=\displaystyle\sqrt{\frac{N_R N_T}{\sum_{i=1}^{N_{\rm cl}}N_{{\rm ray},i}}}$  is a normalization factor ensuring that the received signal power scales linearly with the product $N_R N_T$ \cite{PrecodingHeath}. 
In order to randomly generate the number and the positions of the clusters and of the scatterers, the following assumptions are made. 
The number of clusters\footnote{Although not considered here, it might be reasonable to assume that the number of clusters also depends somehow on the link length, since for increasing distance between the transmit and receive antennas there is a wider space where scatterers might happen to be.}, 
according to \cite{rappaport2014}, is
$N_{\rm cl} \thicksim \text{max}\{{\rm Poisson}(\lambda),1\}$, with $\lambda=1.9$ a suggested value at 73GHz,
and the number of sub-paths for the $i^{th}$ cluster is modeled as a uniform random integer in the range $[1, 30]$ \cite{rappaport2015channel}.
For the generic $i$-th cluster\footnote{At mmWave frequencies the reduced wavelength makes antenna arrays very compact in size, so we can assume that all the antennas in the array see the same scatterers.}, the azimuth departure angles $\phi_{i,l}^t$, $l=1, \ldots, N_{{\rm ray},i}$ are generated according to a  Laplacian distribution\footnote{The Laplacian distribution has been found to be a good fit for a variety of propagation scenarios \cite{spatialcorrelation, PrecodingHeath, 5Gchannelup100GHz}.} whose mean\footnote{Rigorously speaking, the azimuth departure angles 
$\phi_{i,l}^t$ are thus conditionally Laplacian.} $\phi_{i}^t \thicksim {\cal U}[-\pi/2, \pi/2]$, and with standard deviation   $\sigma_{\phi}=5$°. 
A similar reasoning is used for the generation of the elevation departure angles, $\theta_{i,l}^t$ which are indeed assumed to be conditionally Laplacian with a mean $\theta_i^t$ uniformly distributed in $[-\pi/2,  \pi/2]$ and standard deviation $\sigma_{\theta}=5$°. Because of the random orientation of the receiver, also the azimuth and elevation angles of arrival are assumed randomly distributed, again with conditional Laplacian distribution. The  arrival elevation angles, $\theta_{i,l}^r$ are assumed to be Laplacian with mean $\theta_i^r$ uniformly distributed in $[-\pi/2,  \pi/2]$ and standard deviation $\sigma_{\theta}=5$°, while the azimuth arrival angles, $ \phi_{i,l}^r$ are assumed Laplacian with  mean $\phi_i^r$ uniformly distributed in $[0,2 \pi]$ and standard deviation $\sigma_{\phi}=5$° \cite{PrecodingHeath}. For the sake of simplicity, the distance between the transmit antenna array and all the scatterers belonging to the same cluster is constant (i.e. all the scatterers within a cluster are at the same distance from the transmitter), and is generated as a ${\cal U}(1 \, \text{m}, 7/4d)$ random variate, with $d$ the link length; we denote by $r_i$ the distance from the transmitter of the scatterers in the $i$-th cluster. For those clusters whose angle of departure points toward the ground the maximum distance is reduced according to geometrical considerations\footnote{Otherwise stated, we locate on the ground level clusters that would happen to be located underground. }.
The total length $r_{i,l}$ of the $(i,l)$-th propagation path is obtained through geometrical considerations as a function of $d$, and of the heights $h_T$ and $h_R$ of the transmit and receive antenna arrays, respectively\footnote{We consider the single bounce scattering because, according to the vast majority of the existing literature, it is largely predominant at mmWave frequencies.}:
$
r_{i,l}\!=\!r_i\!+\!\sqrt{\!(h_T-h_R\!+r_i\!\sin{\!\theta_{i,l}^t})\!^2\!+\!(d-r_i\cos{\theta_{i,l}^t}\!\cos{\phi_{i,l}^t})^2},
$
for   $l=1, \ldots, N_{{\rm ray}, i}$, and $i=1, \ldots, N_{\rm cl}$.
Regarding the array response vectors $ \mathbf{a}_r(\phi_{i,l}^r,\theta_{i,l}^r)$ and $\mathbf{a}_t(\phi_{i,l}^t,\theta_{i,l}^t)$, a planar antenna array configuration is used for the transmitter and receiver, with $Y_r$, $Z_r$ and $Y_t$, $Z_t$ antennas respectively on the horizontal and vertical axes for the receiver and for the transmitter. Letting $k=2\pi/\lambda$, $\lambda$ the considered wavelength, and denoting by $\tilde{d}$ the inter-element spacing we have 
$$
\begin{array}{lll}
\mathbf{a}_x(\phi_{i,l}^x,\theta_{i,l}^x)=\!\frac{1}{\sqrt{Y_xZ_x}}[1,\ldots,e^{-jk\tilde{d}(m\sin{\phi_{i,l}^x}\sin{\theta_{i,l}^x}+n\cos{\theta_{i,l}^x})},  \\
\ldots,e^{-jk\tilde{d}((Y_x-1)\sin{\phi_{i,l}^x}\sin{\theta_{i,l}^x}+(Z_x-1)\cos{\theta_{i,l}^x})}] \; ,
\end{array}
$$
where $x$ may be either $r$ or $t$.
With regard to the attenuation of the $(i,l)$-th path, we considered the results of  \cite{5Gchannelup100GHz} for four different use-case scenarios: Urban Microcellular (UMi) Open-Square, UMi Street-Canyon, Indoor Hotspot (InH) Office, and InH Shopping Mall. Following \cite{5Gchannelup100GHz},  the attenuation of the  $(i,l)$-th path is written in logarithmic units as
\begin{equation}
L(r_{i,l})\! =\! - 20\log_{10}\!\!\left(\!\frac{4\pi}{\lambda}\!\right) \\ - 10n\! \left[1-b + \! \frac{bc}{\lambda f_0}\right]\! \log_{10}\left(r_{i,l}\right) - X_{\sigma} \; ,
\end{equation}
with 
$n$  the path loss exponent, $X_{\sigma}$  the zero-mean, $\sigma^2$-variance Gaussian-distributed shadow fading term in logarithmic units,  $b$ a system parameter,  and $f_0$  a fixed reference frequency, the centroid of all the frequencies represented by the path loss model. The values for all these parameters for each use-case scenario are reported in Table \ref{PL_parameters}. 
\begin{table}
\caption{Parameters for Path Loss Model \cite{5Gchannelup100GHz}}
\label{PL_parameters}
\centering
\begin{tabular}{|c|c|}
\hline
Scenario & Model Parameters \\
\hline
UMi Street Canyon LOS & $n=1.98 \; , \sigma=3.1 \; \text{dB} \; , b=0$ \\
\hline
UMi Street Canyon NLOS & $n=3.19 \; , \sigma=8.2 \; \text{dB} \; , b=0$ \\
\hline
UMi Open Square LOS & $n=1.85 \; , \sigma=4.2 \; \text{dB} \; , b=0$ \\
\hline
UMi Open Square NLOS & $n=2.89 \; , \sigma=7.1 \; \text{dB} \; , b=0$ \\
\hline
InH Indoor Office LOS & $n=1.73 \; , \sigma=3.02 \; \text{dB} \; , b=0$ \\
\hline
InH Indoor Office NLOS & $n=3.19 \; , \sigma=8.29 \; \text{dB}$ \\ & $ b=0.06 \; , f_0=24.2 \; \text{GHz} $ \\
\hline
InH Shopping Mall LOS & $n=1.73 \; , \sigma=2.01 \; \text{dB} \; , b=0$ \\
\hline
InH Shopping Mall NLOS & $n=2.59 \; , \sigma=7.40 \; \text{dB}$ \\ & $b=0.01 \; , f_0=39.5 \; \text{GHz} $ \\
\hline
\end{tabular}
\end{table}

Let us now comment on the LOS component $\mathbf{H}_{\rm LOS}(\tau)$ in \eqref{eq:channel1}. Denoting by 
$\phi_{\rm LOS}^r$,  $\phi_{\rm LOS}^t$,
$\theta_{\rm LOS}^r$,  and $\theta_{\rm LOS}^t$ the departure angles corresponding to the LOS link, we assume that
\begin{equation}
\begin{array}{llll}
\mathbf{H}_{\rm LOS}(\tau) = &  
I_{\rm LOS}(d) \sqrt{N_R N_T} e^{j \eta} \sqrt{L(d)}\mathbf{a}_r(\phi_{\rm LOS}^r,\theta_{\rm LOS}^r) \times \\ & \mathbf{a}_t^H(\phi_{\rm LOS}^t,\theta_{\rm LOS}^t) h(\tau-\tau_{\rm LOS}) \; .
\end{array}
\label{eq:Hlos}
\end{equation}
In the above equation, $\eta \thicksim \mathcal{U}(0 ,2 \pi)$, while $I_{\rm LOS}(d) $ is a random variate indicating if a LOS link exists between transmitter and receiver. Denoting by $p$ the probability that $I_{\rm LOS}(d) =1$, i.e.,  a LOS link exists, we use again the 
results in  \cite{5Gchannelup100GHz,5G3PPPlikemodel}; for the UMi scenarios, we have:
\begin{equation}
p=\text{min}\left(\frac{20}{d},1\right)\left(1-e^{-\frac{d}{39}}\right)+e^{-\frac{d}{39}} \; ,
\end{equation} 
while for the InH scenarios we have:
\begin{equation}
p=\left\{
\begin{array}{ll}
1 & \; d \leq 1.2  , \\
e^{-\left(\frac{d-1.2}{4.7}\right)} & \;  1.2 <d \leq 6.5 , \\
0.32e^{-\left(\frac{d-6.5}{32.6}\right)} & \; d \geq 6.5.  
\end{array} \right .
\end{equation}
The Matlab script provided in \cite{github_dandrea} permits generating the time-sampled version of the matrix-valued channel in \eqref{eq:channel1}, with both a tunable sampling frequency and shaping filters $h_{\rm TX}(t)$ and $h_{\rm RX}(t)$.

\section{Time-variant channel model}
We now focus on the case in which the channel coherence time is smaller than the observation window, so that the channel cannot be considered as time-invariant. We assume that variations in the channel impulse response are caused by a Doppler frequency induced by the relative motion between transmitter and receiver, and by changes in the scattering coefficients $\alpha_{i,l}$. We assume that both the receiver and the transmitter are moving, with speed $v^{\rm RX}$ and $v^{\rm TX}$ respectively, along the horizontal axis\footnote{This situation may be representative of a device-to-device communication; the extension to the case in which devices do not move along the horizontal axis is neglected for the sake of simplicity.}. The system between section A and section B in Fig. \ref{Fig:Channel} is now modeled as a linear time-variant system whose matrix-valued impulse response is written as
\begin{equation}
\begin{array}{lll}
\mathbf{H}(t,\tau)=\gamma \displaystyle \sum_{i=1}^{N_{\rm cl}}\sum_{l=1}^{N_{{\rm ray},i}}\alpha_{i,l}(t) \sqrt{L(r_{i,l})}   \mathbf{a}_r(\phi_{i,l}^r,\theta_{i,l}^r) \times
\\
 \mathbf{a}_t^H(\phi_{i,l}^t,\theta_{i,l}^t)h(\tau-\tau_{i,l})e^{-j2\pi\nu_{i,l}t} + \mathbf{H}_{\rm LOS}(t,\tau)
\end{array}
\label{eq:channel_LTV}
\end{equation}
Comparing the above equation with the time-invariant channel model in \eqref{eq:channel1}, it is seen that we have introduced Doppler shifts $\nu_{i,l}$ and time-variant scattering 
gains\footnote{Since $v^{\rm RX} \ll c$, $v^{\rm TX} \ll c$ it is easily shown that the arrival and departure angles, the path lengths $r_{i,l}$ and the associated delays $\tau_{i,l}$ can be still considered constant.}. 
The Doppler shifts are expressed as $
\nu_{i,l}=-\frac{f}{c}\left(v^{\rm RX}\cos{\theta_{i,l}^r} \cos{\phi_{i,l}^r}+v^{\rm TX}\cos{\theta_{i,l}^t} \cos{\phi_{i,l}^t}\right) \; ,
$
with $f$ the center frequency of the considered bandwidth.
With regard to the complex path gain, we assume that the discrete-time sequence $\alpha_{i,l}(nT_s)$ that results from the sampling of the received signals with period $T_s$ has an exponential correlation, i.e. 
$
E[\alpha_{i,l}(nT_s) \alpha_{i,l}^*(mT_s)]=\rho^{|m-n|}\; , \forall i, \; \forall l \; .
$
The choice of the coefficient $\rho$ should be a function of both the receiver and the transmitter speeds, $v^{\rm RX}$ and $v^{\rm TX}$, and of the sampling time $T_s$.
The LOS component in \eqref{eq:channel_LTV} is then written as
\begin{equation}
\begin{array}{llll}
\mathbf{H}_{\rm LOS}(t, \tau) = &  
I_{\rm LOS}(d) \sqrt{N_R N_T} e^{j \eta(t)}\sqrt{L(d)}\mathbf{a}_r(\phi_{\rm LOS}^r,\theta_{\rm LOS}^r) \times \\ & \mathbf{a}_t^H(\phi_{\rm LOS}^t,\theta_{\rm LOS}^t) h(\tau-\tau_{\rm LOS})e^{-j 2 \pi \nu_{\rm LOS}t} \; ,
\end{array}
\label{eq:HlosTV}
\end{equation}
with $
\nu_{\rm LOS}\!\!=\!-\frac{f_0}{c}\!\!\left(v^{\rm RX}\!\cos{\theta_{\rm LOS}^r}\!\cos{\phi_{\rm LOS}^r}\!+\!v^{\rm TX}\!\cos{\theta_{\rm LOS}^t}\!\cos{\phi_{\rm LOS}^t}\right),
$ 
and the sampled version of the phase $\eta(t)$ randomly evolves with exponential correlation.
Also for the time-variant channel scenario, the channel generation Matlab script is in \cite{github_dandrea}.


 \section{Simulation Results} 
In this section we show a simple application of the proposed channel model by evaluating the achievable spectral efficiency cumulative distribution function (CDF) for a UMi Street-Canyon scenario, using time-invariant channel model. We consider a single-carrier  transmission of a packet of $L$ data-symbols, with transmit power $P_T=0$ dBW; we denote by  $M$ the multiplexing order, i.e. the number of information symbols that are simultaneously transmitted by the $N_T$ transmit antennas in each symbol interval. 
The transmitter and the receiver are fixed and the heights are $h_T=7$ m and $h_R=1$ m respectively. The inter-element spacing $\tilde{d}$ of the array is assumed to be half-wavelength, and the carrier frequency is $73\,\textrm{GHz}$. Square-root-raised-cosine pulses with roll-off factor 0.22 are adopted as shaping filters both for the transmitter and the receiver. 
The discrete-time baseband equivalent of the received signal at sampling epoch $n$ is written as the $M$-dimensional vector:
\begin{equation}
\mathbf{r}(n)=\sum_{l=0}^{P-1}\mathbf{D}^H \mathbf{H}(l)\mathbf{Q}\mathbf{s}(n-l)+\mathbf{D}^H\mathbf{w}(n)
\end{equation}
Where $P$ is the channel's length, $\mathbf{D}$ is the combining matrix with dimensions $ N_R \times M$, $\mathbf{Q}$ is the precoding matrix with dimensions $ N_T \times M$, $\mathbf{s}(n)$ is the $M$-dimensional data-symbol vector transmitted at time $n$ and $\mathbf{w}(n)$ is the $N_R$-dimensional additive noise vector.
Letting $\mu=\mbox{arg}\max_{\ell = 0, \ldots, \widetilde{P}-1}\left\{ \left\|{\mathbf{H}}(\ell)\right\|_{\rm F}\right\}$,
the matrix $\mathbf{Q}$ contains on its columns the right eigenvectors of the matrix ${\mathbf{H}}(\mu)$ corresponding to its $M$ largest eigenvalues, and the matrix  $\mathbf{D}$ contains on its columns the corresponding left eigenvectors.
In order to have a soft estimate, say $\hat{\mathbf{s}}(n)$, of the data vector $\mathbf{s}(n)$,
we  stack the data-vectors $\mathbf{r}(n), \ldots, \mathbf{r}(n+P-1)$  into the $PM$-dimensional vector $\widetilde{\mathbf{r}}(n)$, and process them through a linear minimum mean square error estimator, represented by the $(MP \times M)$-dimensional matrix $\mathbf{E}$. 
It is easy to show that the LMMSE estimate $\hat{\mathbf{s}}(n)$ can be written as $
\hat{\mathbf{s}}(n)=\mathbf{E}^H\left(\mathbf{A}\mathbf{s}(n) + 
\mathbf{A}_{\rm I}\mathbf{s}_{\rm I}(n) +\mathbf{B}\mathbf{w}
\right) \; ,
$
where $\mathbf{A}$ is an  $(MP \times M)$-dimensional matrix containing the "signatures" of the useful data-symbols, $\mathbf{s}_{\rm I}(n)$ is a data vector containing the interfering symbols falling in the processing window and $\mathbf{A}_{\rm I}$ is the corresponding signatures' matrix. Following \cite{InterferenceMIMO}, the achievable rate in this case is written as:
\begin{equation}
{\cal R}=\log_2 \det \left[\mathbf{I}_M+\mathbf{R}^{-1}\left(\frac{P_T}{M}\mathbf{E}^H\mathbf{A}\mathbf{A}^H\mathbf{E}\right)\right] \; ,
\label{eq:rate}
\end{equation}
with $
\mathbf{R}=\mathbf{E}^H\left(\frac{P_T}{M}\mathbf{A}_{\rm I}\mathbf{A}_{\rm I}^H+ \sigma_N^2\mathbf{B}\mathbf{C_w}\mathbf{B}^H\right)\mathbf{E} $ and  $\mathbf{C_w}=E[\mathbf{w}(n)\mathbf{w}^H(n)]$.  The achievable spectral efficiency is obtained by normalizing \eqref{eq:rate} with respect to the signaling interval and available bandwidth.

\begin{figure}[t]
\centering
\includegraphics[scale=0.25]{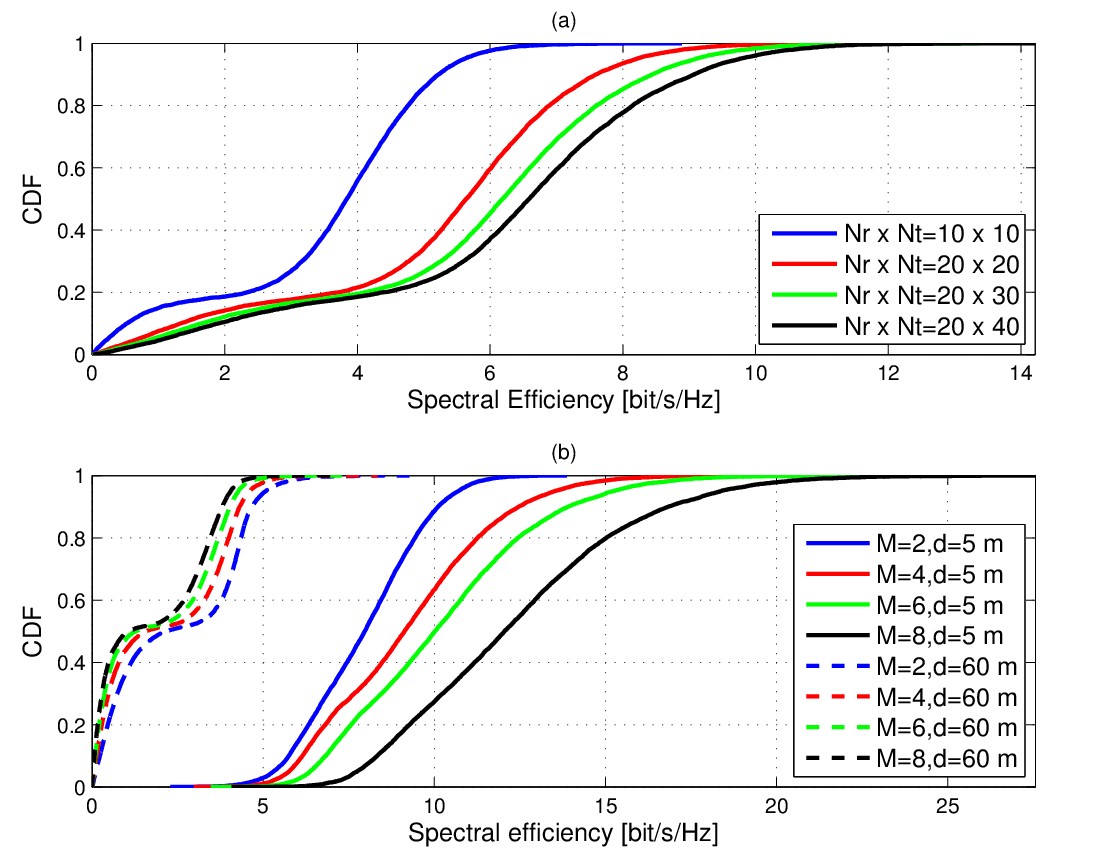}
\caption{(a) shows the CDFs of Spectral Efficiency with $M=4$ , $d=30$ m and varying $N_R \times N_T$; (b) shows the CDFs of Spectral Efficiency for $N_R=20$ and $N_T=30$, for varying $d$ and $M$.}
\label{Fig:fig_subplot}
\end{figure}


Fig. \ref{Fig:fig_subplot} reports the spectral efficiency CDFs for several values of the number of transmit and receive antennas,
of the distance between transmitted and receiver, and of the multiplexing order $M$. It is shown that performance improves with decreasing distance, as well as that increasing the multiplexing order is more beneficial in  the regime of large signal-to-noise ratio (i.e., at shorter distances).

\section{Conclusions}
The paper has provided a clustered  channel model, useful for design and analysis of mmWave MIMO wireless links, along with Matlab scripts for enabling reproducible research.

\bibliography{references}

\bibliographystyle{IEEEtran}

\end{document}